\documentclass[nofootinbib,preprintnumbers,prd,superscriptaddress,twocolumn,showpacs]{revtex4}

\usepackage{amsmath}
\usepackage{amsfonts}
\usepackage{amssymb}
\usepackage{graphicx}
\usepackage[titletoc]{appendix}
\usepackage{color}
\usepackage{hyperref}
\usepackage{cleveref}
\usepackage[rightcaption]{sidecap}
\usepackage{subfigure}
\usepackage{float}
\usepackage{comment}

\usepackage{dcolumn}

\usepackage{array}
\usepackage{ctable}
\usepackage{multirow}
\usepackage{siunitx}
\usepackage{longtable}
\usepackage{tabularx}
\usepackage{booktabs}

\usepackage{amsthm}
\usepackage{mathrsfs}

\usepackage{epsfig}
\usepackage{amscd}

\usepackage{bm}
\usepackage{natbib}
\usepackage{url}
\usepackage{xspace}
\usepackage{kantlipsum}

\graphicspath{{Graphics/}}

\def\be{\begin{equation}}
\def\ee{\end{equation}}
\def\bea{\begin{eqnarray}}
\def\eea{\end{eqnarray}}

\def\nat{Nature}
\def\prl{Phys. Rev. Lett.}

\def\prd{Phys. Rev. D}

\def\mnras{MNRAS}
\def\apj{ApJ}
\def\apjl{ApJ Lett.}

\def\aap{A\&A}

\def\pasj{Publ. Astr. Soc. Japan }
\def\pasp{Publ. Astr. Soc. Pacific}

\def\jcap{JCAP}

\definecolor{vividviolet}{rgb}{0.62, 0.0, 1.0}
\definecolor{amaranth}{rgb}{0.9, 0.17, 0.31}
\definecolor{palatinateblue}{rgb}{0.15, 0.23, 0.89}
\definecolor{brightpink}{rgb}{1.0, 0.0, 0.5}
\definecolor{cornflowerblue}{rgb}{0.39, 0.58, 0.93}
\definecolor{deepcarminepink}{rgb}{0.94, 0.19, 0.22}
\definecolor{radicalred}{rgb}{1.0, 0.21, 0.37}
\hypersetup{ linktoc=all,
    colorlinks, linkcolor={palatinateblue},
    citecolor={brightpink}, urlcolor={amaranth}
}

\begin{document}

\title{Generalizing the relativistic precession model of quasi-periodic oscillations through anharmonic corrections}

\author{Roberto Giambò}
\email{roberto.giambo@unicam.it}
\affiliation{Universit\`a di Camerino, Via Madonna delle Carceri 9, 62032 Camerino, Italy.}
\affiliation{INFN, Sezione di Perugia, Perugia, 06123, Italy.}
\affiliation{INAF - Osservatorio Astronomico di Brera, Milano, Italy.}

\author{Orlando Luongo}
\email{orlando.luongo@unicam.it}
\affiliation{Universit\`a di Camerino, Via Madonna delle Carceri 9, 62032 Camerino, Italy.}
\affiliation{INFN, Sezione di Perugia, Perugia, 06123, Italy.}
\affiliation{INAF - Osservatorio Astronomico di Brera, Milano, Italy.}
\affiliation{SUNY Polytechnic Institute, 13502 Utica, New York, USA.}
\affiliation{Al-Farabi Kazakh National University, Al-Farabi av. 71, 050040 Almaty, Kazakhstan.}

\author{Marco Muccino}
\email{marco.muccino@unicam.it}
\affiliation{Universit\`a di Camerino, Via Madonna delle Carceri 9, 62032 Camerino, Italy.}
\affiliation{NNLOT, Al-Farabi Kazakh National University, Al-Farabi av. 71, 050040 Almaty, Kazakhstan.}
\affiliation{ICRANet, Piazza della Repubblica 10, Pescara, 65122, Italy.}

\author{Alessandro Rossi}
\email{alessandro06.rossi@studenti.unicam.it}
\affiliation{Universit\`a di Camerino, Via Madonna delle Carceri 9, 62032 Camerino, Italy.}

\begin{abstract}
We critically reanalyze the relativistic precession model  of quasi-periodic oscillations, exploring its natural extension beyond the standard harmonic approximation. To do so, we show that the perturbed geodesic equations must include anharmonic contributions arising from the higher-order expansion of the effective potential that cannot be neglected \emph{a priori}, as commonly done in all the approaches pursued so far. More specifically, independently of the underlying spacetime geometry, we find that in the radial sector the non-negligible anharmonic correction is quadratic in the radial displacement, i.e.~$\propto \delta r^2$, and significantly affects the radial epicyclic frequency close to the innermost stable circular orbit. Conversely, polar oscillations $\delta \theta$ remain approximately decoupled from  radial ones, preserving their independent dynamical behavior. To show the need of anharmonic corrections, we thus carry out Monte Carlo–Markov chain analyses on eight neutron star sources of quasi-periodic oscillations. Afterwards, we first work out the outcomes of the harmonic approximation in Schwarzschild, Schwarzschild--de Sitter, and Kerr spacetimes. Subsequently, we apply the anharmonic corrections to them and use it to fit the aforementioned neutron star sources. Our findings indicate that the standard paradigm requires a systematic generalization to include the leading anharmonic corrections that appear physically necessary, although still insufficient to fully account for the observed phenomenology of quasi-periodic oscillations. Accordingly, we speculate on possible refinements of the relativistic precession model, showing the need to revise it at a fundamental level.
\end{abstract}


\maketitle
\tableofcontents

\section{Introduction}

General relativity has been indirectly tested across the last decades through the detection of gravitational waves \cite{2016PhRvL.116f1102A}, black hole (BH) shadows \cite{2019ApJ...875L...1E}, cosmological tensions \cite{Leizerovich:2023qqt}, and so on  \cite{Renevey:2019jrm}. All the checks seemed to validate Einstein's gravity, although possible notable exceptions are found, especially in cosmological realms \cite{Ishak_2018}.

The interest for BHs has, consequently, increased significantly, yielding the need of investigating the collapse conditions more deeply \cite{2011arXiv:1106.5438,2013arXiv:1304.7331,2024arXiv:2412.02740} and to speculate on the nature and origin of regular BHs \cite{Luongo:2025iqq}, where the singularity is removed through a smooth core predicted by the presence of topological charges, vacuum energy, minimal lengths and so on, see e.g. \cite{1968qtr..conf...87B,Bambi2023}.
Nevertheless, the use of BHs in the context of gravitational physics and cosmology has marked the incoming epoch of BH precision astronomy \cite{2021NatRP...3..732V}, where Einstein's theory will be tested in the regime of strong gravity, possibly aiming to show possible regions in which quantum gravity arises \cite{10.1093/mnras/stu1919}.

In this scenario, BHs are not the unique protagonists, i.e., compact objects such as neutron stars (NSs) and white dwarfs (WDs) may represent useful sources of incoming data that, once analyzed, could in principle give us information regarding the main features of these cosmological structures.

Particularly, low-mass X-ray binaries stand up as they exhibit the phenomenon of quasi-periodic oscillations (QPO).
These oscillations manifest as peaks of excess energy from sources such as WDs and/or NSs \cite{2004PASJ...56..553R,2004ApJ...609L..63A,2005A&A...436....1T,2012ApJ...760..138T,2014GrCo...20..233B}, with the remarkable property that this effect may be consequence of mass accretion rate variation.
More precisely, QPO can be set into two main  classes: 1) high frequencies, lying in the range $0.1-1$ kHz, and 2) low frequencies, namely below $0.1$ kHz. Generally, high frequencies are associated with the Keplerian frequency close to the innermost stable circular orbit (ISCO) of a test particle \cite{Lewin_van_der_Klis_2006, Lamb_2008}, whereas low frequencies are determined by the periastron precession frequency \cite{1999NuPhS..69..135S,Boshkayev_2014,Muccino2015,Török_2016,MGM14}.

The frequencies derived from QPO are associated with intrinsic phenomena of compact objects, thus, they represent fundamental tools to Einstein's gravity, comparing spacetime solutions directly with data.
Phrasing it differently, assuming that QPO data at high frequencies resemble the frequencies of a test particle orbiting the compact object close to the ISCO, an examination near this region appears the most likely approach.

Accordingly, there is a wide collection of models aiming to describe the QPO phenomenon \cite{1998ApJ...508..791M, 1998ApJ...499..315T, 2001ApJ...554.1210L, 2001ApJ...553..955F, 2001A&A...374L..19A, 2002ApJ...577L..23T, 2004PASJ...56..553R, 2005PASJ...57L..17K, 2012PASJ...64..139K, 2022JCAP...05..020B}.
Among the \emph{plethora} QPO paradigms, the so-called relativistic precession model (RPM) appears to be the simplest and most consolidate one.
Within this model, the existence of QPO can be justified as a consequence of inhomogeneities of the mass accretion rate of the compact object.

In all the above scenarios, the \emph{Kerr hypothesis} is implicitly assumed. This hypothesis ensures that astrophysical BHs and, more broadly, compact objects, are well-described by the Schwarzschild (S) or Kerr (K) solutions. It is therefore natural to expect that the RPM paradigm supports the validity of this hypothesis, provided that QPO observational data are consistently interpreted within it.

Conversely, this does not appear general, as expected, when adopting the standard treatment of RPM, which exhibits severe statistical inconsistencies, see, e.g., \cite{2023PhRvD.108l4034B}.

Hence, motivated by these issues and within the framework of the standard harmonic approach to RPM, we focus here on including anharmonic terms, derived from the next-order expansion of the effective potential, into the standard RPM paradigm, with the aim of mitigating the unexpected lack of Kerr hypothesis confirmation inside the RPM treatment.

In particular, we show that the resulting coupling terms between radial $\delta r$ and angular $\delta \theta$ oscillations can be neglected, keeping the corresponding equations independent.
Finally, we demonstrate that, under given circumstances, they cannot be neglected \emph{a priori}, albeit insufficient to provide a satisfactory correction to RPM. To show the relevance of higher order terms, we perform Monte Carlo -- Markov chain (MCMC) analyses on QPO data coming from eight NSs. Accordingly, we single out two main hierarchies: in the first one we review the results from S, SdS and Kerr (K) spacetimes within the harmonic approximation, namely the \emph{standard approach} to QPO in the context of RPM \cite{2023PhRvD.108l4034B,2023arXiv230303248}; in the second one, we include the anharmonic contributions and  we consider only K spacetime, showing that anharmonicity is necessary but still provides discrepancies between models and data.  To assess the best-fitting scenarios we critically apply statistical and physical selection criteria.

We conclude that the anharmonic contributions, corresponding to higher-order terms beyond the standard RPM formulation, cannot be neglected as commonly is considered in the literature. In addition, our analysis notably reveals a violation of the Kerr hypothesis, i.e., in order to consistently account for QPOs, the SdS framework remains the most appropriate phenomenological setting, highlighting the limitations of the RPM. This suggests the need of its theoretical extension through additional expansions, further external sources terms and so on. Accordingly, physical consequences based on these aspects are thus developed throughout the manuscript.

The paper is structured as follows. Sect.~\ref{qpo} introduces the QPO phenomenon and the proposed interpretations. Sect.~\ref{epicyclic model} presents the reference models used in Refs.~\cite{2023PhRvD.108l4034B,2023arXiv230303248} and deals with the anharmonic terms coming from the Taylor expansion of the effective potential up to the third order. Sect.~\ref{montecarlo} describes the MCMC analyses performed to establish the best-fitting parameters and models able to describe each QPO source. Section~\ref{final} summarizes the final outcomes, suggesting future perspectives and possible extensions of our work.


\section{Quasi-periodic oscillations}\label{qpo}

QPO are not strictly periodic  and exhibit a slightly variable range of frequencies, each considered roughly constant though.
QPO are  observed in several astrophysical systems, among which,
\begin{itemize}
\item[-] in the power spectra of X-ray binary NSs and BHs, leading to further studies of accretion disks \cite{1985Natur.316..225V, 1986ApJ...306..230M, 2005A&A...440L..25K, 2007PASP..119..393Z, 2012MNRAS.426.1701B, 2014PhRvD..89l7302B},
\item[-] WDs \cite{1982ApJ...257L..71M, 1987A&A...181L..15L, 1989A&A...217..146L,2015A&A...579A..24B}, where QPO were discovered.
\end{itemize}
and, in general, in  strong gravitational fields, induced by the presence of compact objects.

QPO serves as a valuable tool for testing gravitational theories and providing insights about astrophysical sources and cosmology \cite{2009JCAP...09..013B}. Precise measurements of QPO frequencies from accretion disks \cite{Boshkayev:2020kle,Capozziello:2025ycu,Kurmanov:2025uwq,Ashoorioon:2024lso,Kurmanov:2024hpn,Boshkayev:2023fft,Boshkayev:2023ipb,Boshkayev:2022vlv,DAgostino:2022ckg,Kurmanov:2021uqv,Boshkayev:2021chc} around compact objects are crucial for identifying the most accurate models describing the astrophysical processes at work in these systems and, furthermore, in shedding light on the nature of gravity at strong gravity regimes \cite{2011CQGra..28k4009M, 2020EPJC...80..133K}.

The QPO phenomenon and the corresponding physics may be described by several models. Alternative theories of gravity suggest that, through classical methods proper of general relativity, different compact objects are able to generate oscillations \cite{2023PhRvD.108l4034B}. Moreover, it is proven that also BH binary systems are able to yield QPO \cite{arXiv:2111.14329,2022arXiv2206.13609,Tursunov2022,2022arXiv2204.05273}, but in this case the physical origin is unclear, although the correlation with BHs is widely acknowledged.

Among all the treatments under exam, we focus on the widely accepted RPM or \emph{epicyclic model} \cite{1998ApJ...492L..59S,arXiv:astro-ph/9812124,1999ApJ...524L..63S,10.1093/mnras/stac2142,2022arXiv:2107.06828} that makes use of the simplest assumptions, in which test particles, embedded in a strong gravitational field, oscillate as a consequence of the geodesic equation itself.

Very recently, RPM has been tested using QPO data and has provided insights on its validity.
Quite surprisingly, regular BH solutions have been impressively passed the cosmological tests \cite{2023arXiv230303248}, albeit leaving open the problem of having no-hairy terms entering the experimental analysis. Analogously, testing with QPO the goodness of general relativity leads to quite unexpected results, namely, all the numerical analyses showed a clear preference for a SdS spacetime, rather than S or Reissner-Nordstr\"om ones that have been severely ruled out \cite{2023PhRvD.108l4034B}.

Speculation on the lack of data can be easily debated, as well as the need of a vacuum energy term into the metric, associated with a cosmological constant contribution. However, the last possibility appears less likely because compact objects are, in general, quite classical objects, especially if measured from very big distances\footnote{Quantum effects could be found only in proximity of the surface, while the sources are expected to behave classically.}.

\section{The relativistic precession model}\label{epicyclic model}

The harmonic oscillation in question pertains to the fundamental or epicyclic frequencies of test particles traveling in circular orbits within accretion disks around compact objects. To determine these frequencies, we write down the Lagrangian of a test particle with mass $m$
\begin{equation}
    \mathbb L = \frac{1}{2} m g_{\mu\nu} \dot{x}^{\mu} \dot{x}^{\nu}\,,
\end{equation}
where $\dot{x}^{\mu}=d x^{\mu}/d \tau $ is the four-velocity, $g_{\mu\nu}$ is the metric tensor, and $\tau$ is the affine parameter.

Considering a generic stationary, axially-symmetric, and asymptotically-flat metric with line element
\begin{equation}
\label{line}
ds^2 = g_{tt} dt^2 + g_{rr}dr^2 + g_{\theta\theta} d\theta^2
+ 2g_{t\phi}dt d\phi + g_{\phi\phi}d\phi^2 \,,
\end{equation}
the integrals of motion read
\begin{equation}
\dot t = \frac{\mathcal E g_{\phi\phi} + \mathcal L g_{t\phi}}{g_{t\phi}^2 - g_{tt} g_{\phi\phi}},\quad
\dot \phi = - \frac{\mathcal E g_{t\phi} + \mathcal L g_{tt}}{g_{t\phi}^2 - g_{tt} g_{\phi\phi}}\,,
\end{equation}
where the constants of motion $\mathcal E$ and $\mathcal L$ are, respectively, the specific energy and the specific angular momentum.

The normalization condition $g_{\mu\nu} \dot{x}^{\mu} \dot{x}^{\nu}=-1$ of the four velocity of a test-particle leads to
\begin{equation}
\label{norm4vel}
g_{rr} \dot{r}^2 + g_{\theta\theta} \dot{\theta}^2 = \mathcal V\,,
\end{equation}
where we have defined the effective potential
\begin{equation}
 \mathcal V = \frac{\mathcal E^2 g_{\phi\phi} + 2 \mathcal E \mathcal L g_{t\phi} + \mathcal L^2
g_{tt}}{g_{t\phi}^2 - g_{tt} g_{\phi\phi}} - 1  \,.
\end{equation}

\subsection{The ansatz of small oscillations}

To formulate a model of QPO requires that test particles lie on circular orbits in the equatorial plane with
\begin{equation}
\theta_0=\pi/2,\quad\dot{r}=\dot{\theta}=0.
\end{equation}
These conditions are reinforced in the orbital parameters of test particles, namely $\dot t$, $\dot\phi$, and
\begin{subequations}
\begin{align}
\Omega_\phi &= -\frac{\partial_r g_{t\phi} \mp \sqrt{(\partial_r g_{t\phi})^2 - \partial_r g_{tt} \partial_r g_{\phi\phi}}}{\partial_r g_{\phi\phi}}\,,\\
\cal E &= - \frac{g_{tt} + g_{t\phi}\Omega_{\phi}}{
\sqrt{-g_{tt} - 2g_{t\phi}\Omega_{\phi} - g_{\phi\phi}\Omega_{\phi}^2}}\,, \\
\cal L &=  \frac{g_{t\phi} + g_{\phi\phi}\Omega_{\phi}}{
\sqrt{-g_{tt} - 2g_{t\phi}\Omega_{\phi} - g_{\phi\phi}\Omega_{\phi}^2}}\,,
\end{align}
\end{subequations}
where $\pm$ indicates co- and counter-rotating particles, respectively \cite{2016EL....11630006B}, and $\partial_x^n y = \partial^n y/ \partial x^n$ denotes partial derivatives.

The regime of small oscillations holds for a class of geodesics that depart from the equilibrium configurations $x_0=(r_0, \theta_0)$. In particular, the main ansatz is to work with consists in handling the expansion,
\begin{subequations}
    \begin{align}
      r \sim r_0 + \delta r, \\
    \theta \sim \theta_0 + \delta \theta,
    \end{align}
\end{subequations}
giving rise to a set of harmonic differential equations
\begin{equation}\label{harmonic}
\delta r^{\prime\prime} + \Omega^2_r \delta r = 0\,, \qquad
\delta \theta^{\prime\prime} + \Omega^2_\theta \delta \theta = 0,
\end{equation}
where $y^\prime=dy/dt$.
The above differential equations might be computed  in the time domain, $t$. Hence, the corresponding frequencies can be functions of the radial coordinate, $r$, behaving as constants with respect to $t$.
Accordingly, we obtain
\begin{equation}
\label{omegas}
\Omega^2_{r} = - \frac{\partial_r^2 \mathcal V|_{x_0}}{2 g_{rr} \dot t^2},\quad
\Omega^2_{\theta} = - \frac{\partial_\theta^2 \mathcal V|_{x_0}}{2 g_{\theta \theta} \dot t^2}\,,
\end{equation}
that are, in particular, for radial and angular coordinates, respectively.
From these angular frequencies we define the Keplerian frequency and the radial epicyclic frequency of the Keplerian motion, respectively,
\begin{subequations}
\begin{align}
 &f_\phi = \Omega _\phi/(2 \pi),\\
 &f_r = \Omega _r/(2 \pi)\,.
\end{align}
\end{subequations}
Hence, in order to process our model with data within RPM, we make the identification as
\begin{subequations}
\label{QPO_freqs}
\begin{align}
&f_L=f_\phi - f_r,\label{periatron}\\
&f_U = f_\phi,
\end{align}
\end{subequations}
namely, the lower QPO frequency $f_L$ with the periastron precession and the upper QPO frequency $f_U$ with the Keplerian frequency, see Refs. \cite{2014MNRAS.439L..65M,2014MNRAS.437.2554M}.

\subsection{Anharmonic corrections to the precession model}

Assuming displacements $x=(r_0+\delta r, \theta_0+\delta\theta)$, we Taylor expand the effective potential up to the third order:
\begin{align}
\nonumber
\mathcal V (x)\simeq &\,\mathcal V (x_0) + \partial_r \mathcal V|_{x_0}\delta r + \partial_\theta \mathcal V|_{x_0}\delta\theta+\\
\nonumber
&\,\frac{1}{2}\partial^2_r \mathcal V|_{x_0}\delta r^2 + \frac{1}{2}\partial^2_\theta \mathcal V|_{x_0}\delta \theta^2 + \partial_r\partial_\theta \mathcal V|_{x_0}\delta r\delta\theta + \\
\nonumber
&\,\frac{1}{6}\partial^3_r \mathcal V|_{x_0}\delta r^3 + \frac{1}{6}\partial^3_\theta \mathcal V|_{x_0}\delta \theta^3+\\
&\, \frac{1}{2}\partial_r\partial^2_\theta \mathcal V|_{x_0}\delta r\delta\theta^2 + \frac{1}{2}\partial_r^2\partial_\theta \mathcal V|_{x_0}\delta r^2\delta\theta+ \ldots
\end{align}
The stability at the equilibrium points $x_0$ implies
$$\mathcal V(x_0)=\partial_r\mathcal V|_{x_0} = \partial_\theta \mathcal V|_{x_0}=0\,,$$
and since orbits lie in the equatorial plane we get
\begin{equation}
    \partial_r\partial_\theta \mathcal V|_{x_0} = \partial^2_r\partial_\theta \mathcal V|_{x_0} = \partial^3_\theta \mathcal V|_{x_0} = 0.
\end{equation}
Provided that $r^\prime=\delta r^\prime$ and $\theta^\prime =\delta\theta^\prime$, taking the derivative of Eq. \eqref{norm4vel} with respect to $t$ leads to
\begin{align}
\nonumber
& g_{rr}\delta r^\prime \left[\delta r^{\prime\prime} - \frac{\partial_r^2 \mathcal V|_{x_0}}{2 g_{rr} \dot t^2} \delta r - \frac{\partial^3_r \mathcal V|_{x_0}}{4g_{rr}\dot t^2}\delta r^2 - \frac{\partial_r\partial_\theta^2 \mathcal V|_{x_0}}{4g_{rr}\dot t^2}\delta\theta^2\right] +\\
& g_{\theta\theta}\delta\theta^\prime \left[\delta\theta^{\prime\prime} -  \frac{\partial_\theta^2 \mathcal V|_{x_0}}{2g_{\theta\theta}\dot t^2}\delta\theta - \frac{\partial_r\partial_\theta^2 \mathcal V|_{x_0}}{2g_{\theta\theta}\dot t^2}\delta\theta \delta r\right]=0\,.
\end{align}
Thus, the above equation holds true by solving the following two ordinal differential equations (ODEs)
\begin{subequations}
\label{ODE}
\begin{align}
&\delta r^{\prime\prime} + \alpha_0 \delta r + \alpha_1\delta r^2 + \alpha_2 \delta\theta^2=0\,,\label{mar}\\
&\delta \theta^{\prime\prime} +\beta_0 \delta\theta + \beta_1\delta\theta\delta r=0\,,\label{yam}
\end{align}
\end{subequations}
where we have defined
\begin{align}
&\alpha_0=-\frac{\partial_r^2\nonumber \mathcal V|_{x_0}}{2 g_{rr} \dot t^2}
\ ,\
\alpha_1=-\frac{\partial^3_r \mathcal V|_{x_0}}{4g_{rr}\dot t^2}
\ ,\
\alpha_2 = - \frac{\partial_r\partial_\theta^2 \mathcal V|_{x_0}}{4g_{rr}\dot t^2},\\
\nonumber
&\beta_0 = -\frac{\partial_\theta^2 \mathcal V|_{x_0}}{2g_{\theta\theta}\dot t^2}
\ ,\
\beta_1 = -\frac{\partial_r\partial_\theta^2 \mathcal V|_{x_0}}{2g_{\theta\theta}\dot t^2}\,.
\end{align}

Eqs.~\eqref{ODE} represent coupled ODEs.
To evaluate the magnitude of harmonic and anharmonic terms in both radial $\delta r$ and polar $\delta\theta$ oscillations, we consider the specific case of K spacetime
\begin{subequations}
\label{gKerr}
\begin{align}
g_{tt} = & - \left(1 - \frac{2 M r}{\Sigma}\right),\quad g_{rr} = \frac{\Sigma}{\Delta}\,,\\
g_{\theta\theta} = &\,\Sigma\quad,\quad g_{t\phi} = - \frac{2 a M r}{\Sigma} \sin^2\theta\,,\\
g_{\phi\phi} = & \left(r^2 + a^2
+ \frac{2 a^2 M r \sin^2\theta}{\Sigma} \right) \sin^2\theta\,,
\end{align}
\end{subequations}
where we have defined the specific angular momentum $a = j M$, the mass $M$ of the central object, the spin parameter $j$, and the auxiliary functions $\Sigma = r^2 + a^2 \cos^2\theta$ and
$\Delta = r^2 - 2 M r + a^2$. The Schwarzschild (S) spacetime is recovered for $j\rightarrow0$ (or $a\rightarrow0$).

\begin{figure*}
{\includegraphics[width=0.49\hsize,clip]{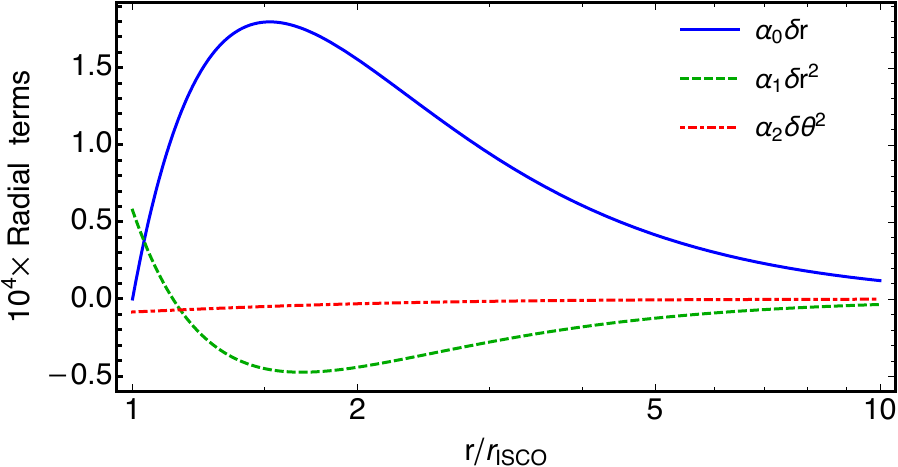} \hfill
\includegraphics[width=0.48\hsize,clip]{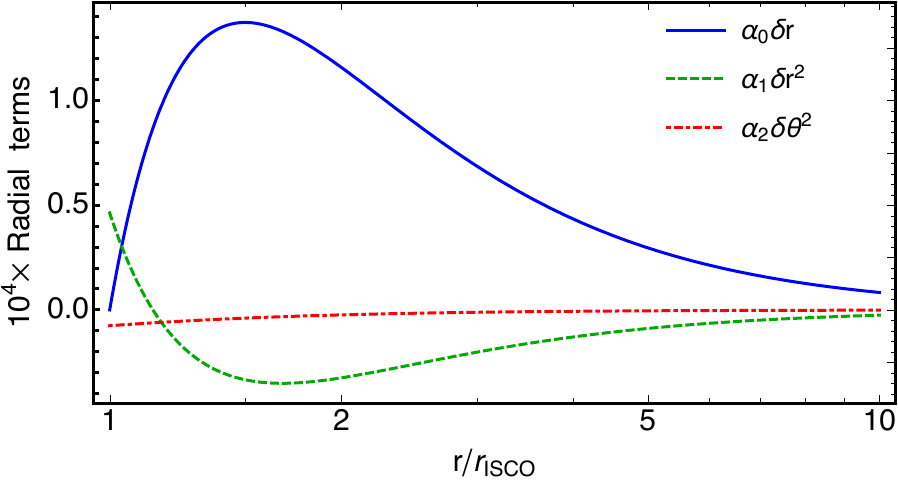}}\\
{\includegraphics[width=0.49\hsize,clip]{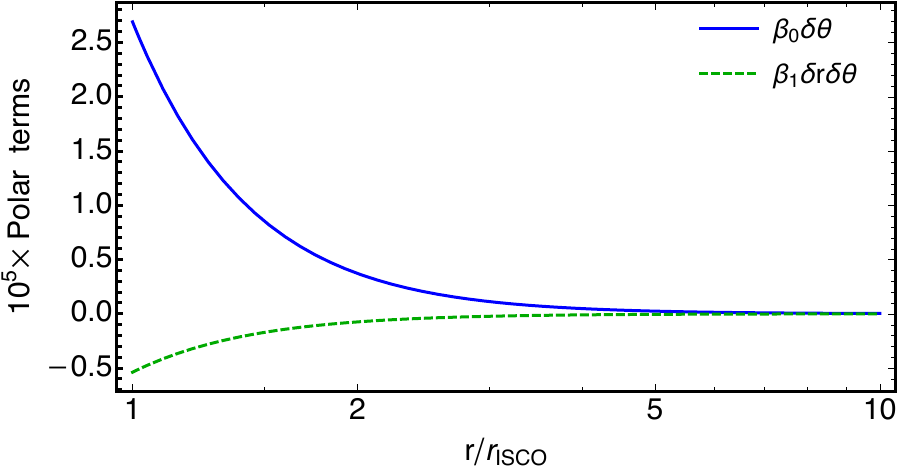}\hfill
\includegraphics[width=0.48\hsize,clip]{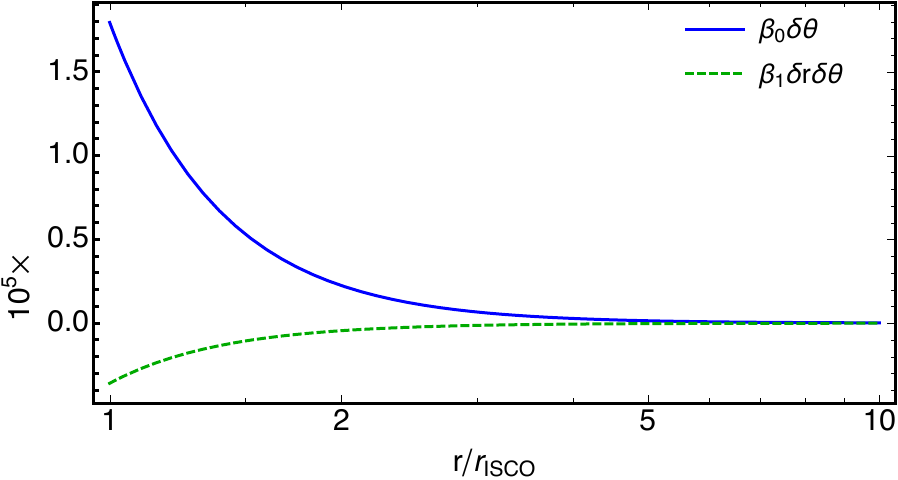}}
\caption{\emph{Left panels}: radial (top) and polar (bottom) harmonic and anharmonic terms for K spacetime with $M=3$ km, $j=0.3$ (co-rotation), $\delta r=0.1r$, and $\delta\theta=2^{\circ}$. \emph{Right panels}: same terms as before and same parameters $M$, $\delta r$, and $\delta \theta$ for S spacetime.}
\label{fig:ODEs}
\end{figure*}

Clearly, the harmonic terms ($\alpha_0\delta r$ and $\beta_0 \delta\theta$) and anharmonic ones ($\alpha_1\delta r^2$, $\alpha_2\delta\theta^2$, and $\beta_1 \delta r\delta\theta$) depend upon the radial distance $r$ of test particles with respect to the central object, which cannot be smaller than the ISCO of the considered spacetime, i.e., $r>r_{\rm ISCO}$.
In this respect, it is particular helpful to visualize the above terms as functions of the ratio $r/r_{\rm ISCO}$ defining \citep{1972ApJ...178..347B}
\begin{subequations}
\label{eq:riscoKerr}
\begin{align}
r_{\rm ISCO}&=M\left[3+z_2\mp \sqrt{(3-z_1)(3+z_1+2z_2)}\right]\,,\\
z_1&= 1+\sqrt[3]{1-j^2}\left(\sqrt[3]{1+j}+\sqrt[3]{1-j}\right)\,,\\
z_2&= \sqrt{3 j^2+z^2_1}\,,
\end{align}
\end{subequations}
where $\mp$ indicates co- and counter-rotation, respectively.

Figure~\ref{fig:ODEs} portraits the behaviors of harmonic and anharmonic terms of Eqs.~\eqref{ODE} as functions of $r/r_{\rm ISCO}$ for co-rotating particles around a K spacetime with $M=3$ km and $j=0.3$ (left plots) and a S spacetime with $M=3$ km (right plots). In both the cases, we assumed $\delta r=0.1r$ and $\delta\theta=2^{\circ}$, for the sake of a clear visualization.
Similar behaviors can be obtained for counter-rotating particles.

From Figure~\ref{fig:ODEs} we conclude that:
\begin{enumerate}
\item[-] the harmonic terms $\alpha_0\delta r$ and $\beta_0\delta\theta$ are the dominant contributions at almost every $r$;
\item[-] the anharmonic term $\alpha_1\delta r^2$ dominates over $\alpha_0\delta r$ at $r\approx r_{\rm ISCO}$, whereas at larger $r$ its magnitude is $\approx20\%$--$30\%$ of $\alpha_0\delta r$;
\item[-] the anharmonic term $\alpha_2\delta\theta^2$ is always negligible with respect to $\alpha_0\delta r$ and $\alpha_1\delta r^2$ at every $r$;
\item[-] the anharmonic term $\beta_1\delta r \delta\theta$ is $\approx20\%$ of $\beta_0\delta\theta$ at every $r$, thus it cannot be neglected.
\end{enumerate}

The above considerations enable us to remove the couplings between the two ODEs.
Focusing on the radial oscillations, that contribute to the periastron precession frequency in Eq.~\eqref{periatron}, only the $\delta r^2$ correction remains and Eq.~\eqref{mar} can be written as\footnote{The anharmonic terms coming from the fourth order Taylor expansion of the effective potential are negligible.}
\begin{equation}
\delta r^{\prime\prime} + \alpha_0 \delta r + \alpha_1\delta r^2 =0\,,\label{mar2}
\end{equation}
that is the ODE of an undamped Helmholtz oscillator.

\subsection{Anharmonicity vs radial epyciclic frequency}

Following Ref.~\cite{HU2006462}, the solution of Eq.~\eqref{mar2} is
\begin{equation}
\delta r(t) = A + B\,{\rm sn}^2(\omega t, k)\,,\label{sinJacobi}
\end{equation}
where ${\rm sn}(u,k)$ is the Jacobi elliptic sine function with elliptic integral of the first kind $u$ and elliptic modulus $k$.
We insert Eq.~\eqref{sinJacobi} into Eq.~\eqref{mar2} and obtain
\begin{align}
\nonumber
&\left(\alpha_0 A + \alpha_1 A^2 + 2 B \omega^2\right) +\\
\nonumber
& \left(\alpha_0 B + 2 \alpha_1 A B - 4 B \omega^2 - 4 B k \omega^2 \right) {\rm sn}^2(\omega t, k) +\\
&\left(\alpha_1 B^2 + 6 B k \omega^2 \right) {\rm sn}^4(\omega t, k)=0\,.
 \end{align}
Next, imposing the coefficients of the ${\rm sn}^n(u,k)$ terms to vanish, we obtain the following conditions
\begin{subequations}
\label{cond2}
\begin{align}
B&=-\frac{6 k \omega^2}{\alpha_1},\\
\omega&=\frac{1}{2}\sqrt{\frac{\alpha_0 + 2 \epsilon}{1+k}},\\
k&=\frac{1}{2}\!+\!\frac{3 \alpha_0^2 \pm (\alpha_0 + 2 \epsilon)\sqrt{3(\alpha_0 - 2\epsilon)(3\alpha_0 + 2\epsilon)}}{8 \epsilon (\alpha_0 + \epsilon)},
\end{align}
\end{subequations}
where $\epsilon=\alpha_1 A$. Requiring that for $\alpha_1\rightarrow0$ (or $\epsilon\rightarrow0$) the solution in Eq.~\eqref{sinJacobi} reduces to the harmonic solution $\delta r(t) = A \cos(\sqrt{\alpha_0} t)$, namely when $k=0$, $\omega=\sqrt{\alpha_0}/2$, and $B=-2A$, Eqs.~\eqref{cond2} can be rewritten as
\begin{subequations}
\label{cond3}
\begin{align}
B&=-\frac{12A(\alpha_0+\epsilon)}{3(\alpha_0+2\epsilon) \sqrt{3(\alpha_0 - 2\epsilon)(3\alpha_0 + 2\epsilon)}},\\
\label{cond3o}
\omega&=\frac{1}{2}\sqrt{\frac{1}{2} (\alpha_0 + 2\epsilon) + \frac{1}{6} \sqrt{3(\alpha_0 - 2\epsilon) (3\alpha_0 + 2\epsilon)}},\\
k&=\frac{1}{2}+\frac{3(2\epsilon^2 + 2\alpha_0\epsilon -\alpha_0^2)}{3 \alpha_0^2 + (\alpha_0 + 2 \epsilon)\sqrt{3(\alpha_0 - 2\epsilon)(3\alpha_0 + 2\epsilon)}}.
\end{align}
\end{subequations}

Knowing Eq.~\eqref{cond3o} and the properties of the function ${\rm sn}(u,k)$, we can compute the exact period $T$ and the angular frequency $\overline\Omega_r$ of the oscillations as
\begin{equation}
T=\frac{2I}{\omega},\qquad \overline\Omega_r=\frac{\pi\omega}{I},
\end{equation}
where $I$ is the complete elliptical integral of the first kind
\begin{equation}
I=\int^{\pi/2}_{0}{\frac{d\xi}{\sqrt{1-k\sin^2\xi}}}\,,
\end{equation}
and for $k=0$ is easy to show that $I=\pi/2$, $T=2\pi/\sqrt{\alpha_0}$, and $\overline\Omega_r=\Omega_r=\sqrt{\alpha_0}$.

The expression of the \emph{generalized radial epicyclic angular frequency} $\overline{\Omega}_r$ is very complicated, but it can display its main features by expanding it\footnote{This expansion only provides a clear visualization of the anharmonic correction. In the following, we will use the full correction.} around $\alpha_1\approx0$
\begin{equation}
\overline{\Omega}_r=\sqrt{\alpha_0}\left(1 - \frac{5 A^2 \alpha_1^2}{12\alpha_0^2}\right)+\mathcal O(\alpha_1^3)\,,
\end{equation}
where \cite{1972ApJ...178..347B}
\begin{subequations}
\begin{align}
\alpha_0 =&\,\Omega_\phi^2\left(1-\frac{6M}{r} \pm 8a\sqrt{\frac{M}{r^3}} - \frac{3a^2}{r^2}\right),\\
\nonumber
\alpha_1 =&\, \Omega_\phi^2\left\{\frac{9}{r} - \frac{6M}{r^2} \mp 6 a \sqrt{\frac{M}{r^5}} + \frac{3a^2}{r^3} +\right.\\
&\,\left.12\left(1-\frac{M}{r}\right)\left[\frac{3M -r \mp 2a \sqrt{M/r}}{a^2+r (r-2 M)}\right]\right\},\\
\Omega_\phi =&\, \frac{\pm  M}{\sqrt{Mr^3}\pm a M}\,,
\end{align}
\end{subequations}
and upper and lower signs refer to co- and counter-rotating particles, respectively.

Unlike the harmonic case, now the radial angular frequency depends upon the radial oscillation amplitude $A$. Specifically, $A$ is a constant at a fixed $r_0$, but more generally it is a function of $r$, namely $A(r)$. Hereby, we consider three functions (1) constant $A_0$, (2) Poissonian $A_0 \sqrt{r}$, and (3) logarithmic $A_0 \ln r$, and the corresponding anharmonic cases are labeled as $A_1$, $A_2$ and $A_3$.

Finally, the \emph{anharmonic Keplerian frequencies} are
\begin{subequations}
    \begin{align}
 &f_\phi = \Omega _\phi/(2 \pi),\\
 &f_r = \overline{\Omega}_r/(2 \pi),
    \end{align}
\end{subequations}
and, combined as in Eqs.~\eqref{QPO_freqs}, can be used to fit the data.

\section{Statistical analysis}\label{montecarlo}

We compare our theoretical findings directly with QPO data.
To do so we follow this strategy based on
\begin{itemize}
\item[-] fitting data with the anharmonic correction, and
\item[-] statistically comparing these results with the harmonic case from S, SdS, and K spacetimes.
\end{itemize}

Clearly, the results are functions of the free parameters of the underlying spacetimes, adopted for our purposes.
Based on Metropolis-Hastings algorithm MCMC analyses, the anharmonic  best-fit parameters $p$ are deduced by maximizing the log-likelihood (LLH) function
\begin{equation}
\label{loglike}
    \ln L = -\sum_{k=1}^{N}\left\{\dfrac{\left[f_{\rm L}^k-f_{\rm L}(p,f_{\rm U}^k)\right]^2}{2(\sigma f_{\rm L}^k)^2} + \ln(\sqrt{2\pi}\sigma f_{\rm L}^k)\right\},
\end{equation}
applied to $N$ data for each source, sampled as lower frequencies $f_{\rm L}^k$, their errors $\sigma f_{\rm L}^k$, and upper frequencies $f_{\rm U}^k$.

With reference to the line element Eq.~\eqref{line}, for the harmonic case we considered K spacetime with metric tensor coefficients given by  Eqs.~\eqref{gKerr}, whereas for SdS spacetime the following metric tensor coefficients are
\begin{subequations}
\label{gSdS}
\begin{align}
g_{tt} = & - g_{rr}^{-1} = - \left(1 - \frac{2 M}{r} - \frac{R_0}{12}r^2\right),\\
g_{\theta\theta} = &\, r^2,\quad g_{t\phi} = 0,\quad g_{\phi\phi} = \, r^2 \sin^2\theta\,.
\end{align}
\end{subequations}
It is trivial to notice that S spacetime is recovered by imposing $a=0$ in Eqs.~\eqref{gKerr}, or $R_0=0$ in Eqs.~\eqref{gSdS}.

In the harmonic case, the best-fit parameters of S and SdS spacetimes were obtained in Ref.~\cite{2023PhRvD.108l4034B}, whereas the corresponding ones for K spacetime can be found in Ref.~\cite{2024MNRAS.531.3876B}. In all cases, the best-fit results (see Table~\ref{tab:results_MCMC}) were obtained employing the same kind of MCMC analyses and QPO sources utilized in this work.

\begin{table*}
\centering
\scriptsize
\setlength{\tabcolsep}{0.7em}
\renewcommand{\arraystretch}{1.3}
\begin{tabular}{lllccccrrrrr}
\hline\hline
Source                                  &
Case                                    &
Metric                                  &
$M\,({\rm M}_\odot)$                    &
$R_0\,(10^{-5}{\rm km}^{-2})$           &
$j$                                     &
$A_0\,({\rm km})$                       &
$\ln L_0$                               &
AIC                                     &
BIC                                     &
$\Delta$AIC                             &
$\Delta$BIC                             \\
\hline
Cir X1                                  &
H                                       &
S                                       &
$2.224^{+0.029}_{-0.029}$               &
--                                      &
--                                      &
--                                      &
$-126$ & $254$ & $254$ & $110$ & $109$  \\
                                        &
H                                       &
SdS                                     &
$1.846^{+0.045}_{-0.045}$               &
$1.28^{+0.12}_{-0.12}$                  &
--                                      &
--                                      &
$-70$ & $144$ & $145$ & $0$ & $0$  \\
                                        &
H                                       &
K                                       &
$5.12^{+0.15}_{-0.24}$                  &
--                                      &
$0.995^{+0.005}_{-0.028}$               &
--                                      &
$-77$ & $159$ & $160$ & $15$ & $15$  \\
                                        &
A$_1$                                   &
K                                       &
$5.01^{+0.24}_{-0.15}$                  &
--                                      &
$0.994^{+0.006}_{-0.028}$               &
$0.50^{+0.46}_{-0.48}$                  &
$-77$ & $161$ & $162$ & $17$ & $17$    \\
                                        &
A$_2$                                   &
K                                       &
$4.40^{+0.83}_{-0.88}$                  &
--                                      &
$0.993^{+0.007}_{-0.043}$               &
$0.75^{+0.26}_{-0.54}$                  &
$-77$ & $161$ & $162$ & $17$ & $17$    \\
                                        &
A$_3$                                   &
K                                       &
$4.96^{+0.27}_{-0.21}$                  &
--                                      &
$0.995^{+0.005}_{-0.036}$               &
$0.23^{+0.45}_{-0.23}$                  &
$-77$ & $161$ & $162$ & $17$ & $17$    \\
\hline
GX 5-1                                  &
H                                       &
S                                       &
$2.161^{+0.010}_{-0.010}$               &
--                                      &
--                                      &
--                                      &
$-200$ & $403$ & $404$ & $187$ & $186$  \\
                                        &
H                                       &
SdS                                     &
$2.397^{+0.019}_{-0.019}$               &
$-6.46^{+0.48}_{-0.48}$                 &
--                                      &
--                                      &
$-106$ & $216$ & $218$ & $0$ & $0$\\
                                        &
H                                       &
K                                       &
$1.286^{+0.027}_{-0.014}$               &
--                                      &
$-0.986^{+0.048}_{-0.014}$              &
--                                      &
$-142$ & $287$ & $289$ & $71$ & $71$   \\
                                        &
A$_1$                                   &
K                                       &
$1.288^{+0.023}_{-0.017}$               &
--                                      &
$-0.987^{+0.047}_{-0.013}$              &
$0.23^{+0.45}_{-0.23}$                  &
$-142$ & $289$ & $292$ & $73$ & $74$    \\
                                        &
A$_2$                                   &
K                                       &
$1.288^{+0.025}_{-0.016}$               &
--                                      &
$-0.991^{+0.052}_{-0.009}$              &
$0.039^{+0.095}_{-0.039}$               &
$-142$ & $289$ & $292$ & $73$ & $74$    \\
                                        &
A$_3$                                   &
K                                       &
$1.288^{+0.025}_{-0.017}$               &
--                                      &
$-0.991^{+0.053}_{-0.009}$              &
$0.070^{+0.138}_{-0.070}$               &
$-142$ & $289$ & $292$ & $73$ & $74$    \\
\hline
GX 17+2                                 &
H                                       &
S                                       &
$2.0768^{+0.0002}_{-0.0003}$            &
--                                      &
--                                      &
--                                      &
$-1819$ & $3640$ & $3641$ & $3543$ & $3543$  \\
                                        &
H                                       &
SdS                                     &
$1.733^{+0.011}_{-0.011}$               &
$21.53^{+0.45}_{-0.45}$                 &
--                                      &
--                                      &
$-46$ & $97$ & $98$ & $0$ & $0$ \\
                                        &
H                                       &
K                                       &
$8.60^{+0.14}_{-0.16}$                  &
--                                      &
$0.9961^{+0.004}_{-0.005}$              &
--                                      &
$-53$ & $111$ & $112$ & $14$ & $14$   \\
                                        &
A$_1$                                   &
K                                       &
$2.827^{+0.088}_{-0.108}$               &
--                                      &
$0.270^{+0.035}_{-0.047}$               &
$5.05^{+0.39}_{-0.36}$                  &
$-184$ & $374$ & $376$ & $277$ & $278$    \\
                                        &
A$_2$                                   &
K                                       &
$1.415^{+0.101}_{-0.044}$               &
--                                      &
$-0.934^{+0.142}_{-0.064}$              &
$9.18^{+0.28}_{-0.17}$                  &
$-350$ & $706$ & $708$ & $609$ & $610$    \\
                                        &
A$_3$                                   &
K                                       &
$1.433^{+0.142}_{-0.052}$               &
--                                      &
$-0.907^{+0.188}_{-0.078}$              &
$13.45^{+0.37}_{-0.32}$                 &
$-359$ & $724$ & $726$ & $627$ & $628$    \\
\hline
GX 340+0                                &
H                                       &
S                                       &
$2.1023^{+0.0033}_{-0.0033}$            &
--                                      &
--                                      &
--                                      &
$-131$ & $264$ & $264$ & $13$ & $12$  \\
                                        &
H                                       &
SdS                                     &
$2.149^{+0.015}_{-0.015}$               &
$-1.39^{+0.45}_{-0.44}$                 &
--                                      &
--                                      &
$-126$ & $256$ & $257$ & $5$ & $5$  \\
                                        &
H                                       &
K                                       &
$1.56^{+0.16}_{-0.12}$                  &
--                                      &
$-0.52^{+0.18}_{-0.17}$                 &
--                                      &
$-125$ & $254$ & $255$ & $3$ & $3$   \\
                                        &
A$_1$                                   &
K                                       &
$1.35^{+0.13}_{-0.08}$                  &
--                                      &
$-0.80^{+0.18}_{-0.15}$                 &
$1.66^{+0.80}_{-1.28}$                  &
$-123$ & $252$ & $253$ & $1$ & $1$    \\
                                        &
A$_2$                                   &
K                                       &
$5.17^{+0.46}_{-0.21}$                  &
--                                      &
$-0.89^{+0.22}_{-0.11}$                 &
$2.116^{+0.015}_{-0.031}$               &
$-122$ & $251$ & $252$ & $0$ & $0$    \\
                                        &
A$_3$                                   &
K                                       &
$4.75^{+0.33}_{-0.19}$                  &
--                                      &
$-0.92^{+0.15}_{-0.08}$                 &
$3.556^{+0.017}_{-0.023}$               &
$-124$ & $254$ & $255$ & $3$ & $3$    \\
\hline
Sco X1                                  &
H                                       &
S                                       &
$1.9649^{+0.0005}_{-0.0005}$            &
--                                      &
--                                      &
--                                      &
$-3887$ & $7776$ & $7778$ & $7509$ & $7507$  \\
                                        &
H                                       &
SdS                                     &
$1.690^{+0.003}_{-0.003}$               &
$21.77^{+0.24}_{-0.25}$                 &
--                                      &
--                                      &
$-159$ & $321$ & $325$ & $54$ & $54$  \\
                                        &
H                                       &
K                                       &
$6.352^{+0.062}_{-0.068}$               &
--                                      &
$0.9274^{+0.0033}_{-0.0037}$            &
--                                      &
$-132$ & $267$ & $271$ & $0$ & $0$   \\
                                        &
A$_1$                                   &
K                                       &
$5.98^{+0.30}_{-0.27}$                  &
--                                      &
$0.910^{+0.015}_{-0.018}$              &
$2.14^{+0.59}_{-1.40}$                  &
$-132$ & $269$ & $275$ & $2$ & $4$    \\
                                        &
A$_2$                                   &
K                                       &
$5.96^{+0.34}_{-0.31}$                  &
--                                      &
$0.915^{+0.010}_{-0.027}$               &
$0.39^{+0.18}_{-0.24}$                  &
$-132$ & $269$ & $275$ & $2$ & $4$    \\
                                        &
A$_3$                                   &
K                                       &
$5.78^{+0.53}_{-0.35}$                  &
--                                      &
$0.898^{+0.028}_{-0.026}$               &
$0.79^{+0.22}_{-0.39}$                  &
$-132$ & $269$ & $275$ & $2$ & $4$    \\
\hline
4U1608–52                               &
H                                       &
S                                       &
$1.9596^{+0.0038}_{-0.0038}$            &
--                                      &
--                                      &
--                                      &
$-236$ & $474$ & $474$ & $350$ & $349$  \\
                                        &
H                                       &
SdS                                     &
$1.728^{+0.014}_{-0.014}$               &
$17.62^{+0.94}_{-0.94}$                 &
--                                      &
--                                      &
$-66$ & $136$ & $137$ & $12$ & $12$ \\
                                        &
H                                       &
K                                       &
$5.94^{+0.30}_{-0.26}$                  &
--                                      &
$0.906^{+0.019}_{-0.018}$               &
--                                      &
$-61$ & $126$ & $127$ & $2$ & $2$   \\
                                        &
A$_1$                                   &
K                                       &
$4.46^{+0.40}_{-0.26}$                  &
--                                      &
$0.782^{+0.043}_{-0.032}$               &
$4.19^{+0.81}_{-1.44}$                  &
$-59$ & $124$ & $125$ & $0$ & $0$    \\
                                        &
A$_2$                                   &
K                                       &
$4.62^{+0.97}_{-0.21}$                  &
--                                      &
$0.854^{+0.031}_{-0.079}$               &
$0.65^{+0.30}_{-0.38}$                  &
$-60$ & $126$ & $127$ & $2$ & $2$    \\
                                        &
A$_3$                                   &
K                                       &
$4.52^{+0.48}_{-0.26}$                  &
--                                      &
$0.790^{+0.048}_{-0.033}$               &
$1.16^{+0.34}_{-0.32}$                  &
$-59$ & $125$ & $126$ & $1$ & $1$    \\
\hline
4U1728–34                               &
H                                       &
S                                       &
$1.7345^{+0.0028}_{-0.0029}$            &
--                                      &
--                                      &
--                                      &
$-213$ & $427$ & $427$ & $353$ & $353$  \\
                                        &
H                                       &
SdS                                     &
$1.445^{+0.016}_{-0.016}$               &
$30.74^{+1.58}_{-1.58}$                 &
--                                      &
--                                      &
$-35$ & $74$ & $74$ & $0$ & $0$ \\
                                        &
H                                       &
K                                       &
$6.57^{+0.30}_{-0.32}$                  &
--                                      &
$0.981^{+0.011}_{-0.014}$               &
--                                      &
$-36$ & $76$ & $76$ & $2$ & $2$   \\
                                        &
A$_1$                                   &
K                                       &
$3.28^{+0.71}_{-0.25}$                  &
--                                      &
$0.681^{+0.128}_{-0.061}$               &
$5.38^{+0.69}_{-0.67}$                  &
$-35$ & $75$ & $75$ & $1$ & $1$    \\
                                        &
A$_2$                                   &
K                                       &
$2.987^{+0.067}_{-0.105}$               &
--                                      &
$0.877^{+0.023}_{-0.025}$               &
$1.303^{+0.015}_{-0.017}$               &
$-36$ & $79$ & $79$ & $5$ & $5$    \\
                                        &
A$_3$                                   &
K                                       &
$3.026^{+0.095}_{-0.344}$               &
--                                      &
$-0.845^{+0.030}_{-0.041}$              &
$1.971^{+0.030}_{-0.041}$               &
$-37$ & $81$ & $81$ & $7$ & $7$    \\
\hline
4U0614+091                              &
H                                       &
S                                       &
$1.9042^{+0.0014}_{-0.0014}$            &
--                                      &
--                                      &
--                                      &
$-843$ & $1688$ & $1689$ & $1411$ & $1409$  \\
                                        &
H                                       &
SdS                                     &
$1.545^{+0.011}_{-0.011}$               &
$28.39^{+0.80}_{-0.80}$                 &
--                                      &
--                                      &
$-189$ & $382$ & $384$ & $105$ & $104$ \\
                                        &
H                                       &
K                                       &
$7.708^{+0.043}_{-0.139}$               &
--                                      &
$0.9989^{+0.0011}_{-0.0042}$            &
--                                      &
$-155$ & $315$ & $317$ & $38$ & $37$   \\
                                        &
A$_1$                                   &
K                                       &
$2.72^{+0.37}_{-1.00}$                  &
--                                      &
$0.694^{+0.058}_{-0.308}$               &
$6.771^{+0.063}_{-0.184}$               &
$-135$ & $277$ & $280$ & $0$ & $0$    \\
                                        &
A$_2$                                   &
K                                       &
$1.18^{+0.58}_{-0.05}$                  &
--                                      &
$0.505^{+0.059}_{-0.297}$               &
$1.439^{+0.008}_{-0.027}$               &
$-140$ & $286$ & $289$ & $9$ & $9$    \\
                                        &
A$_3$                                   &
K                                       &
$1.04^{+0.12}_{-0.20}$                  &
--                                      &
$0.03^{+0.12}_{-0.27}$                  &
$2.072^{+0.026}_{-0.061}$               &
$-136$ & $279$ & $282$ & $2$ & $2$    \\
\hline
\end{tabular}
\caption{MCMC results in the harmonic case H for S, SdS, and K spacetimes, and in the anharmonic cases A$_1$, A$_2$ and A$_3$, with $A(r)=\{A_0,A_0\sqrt{r},A_0\ln r\}$, respectively, for K metric.}
\label{tab:results_MCMC}
\end{table*}

For the anharmonic fits we consider as priors
$$
M\in [0,10]\,{\rm M}_\odot,\quad j\in [-1,1],\quad A_0\in[0,15]\,{\rm km}\,,
$$
which are large enough to get the absolute maximum $\ln L_0$ of Eq.~\eqref{loglike}, allowing constraints that may contrast with the NS interpretation.
For each source
\begin{itemize}
\item[-] the LLH is computed from $\mathcal N \simeq 10^5$ iterations, and
\item[-] the best-fit parameters are determined from $\ln L_0$.
\end{itemize}

\subsection{Statistical selection criteria}

Albeit each QPO source has significantly different statistical response, set of data points $N$, and given number $p$ of free parameters, from the overall analysis we can assess the best-fit model by using the Aikake Information Criterion (AIC) and the Bayesian Information Criterion (BIC) \cite{2007MNRAS.377L..74L}, respectively defined as
\begin{subequations}
\begin{align}
{\rm AIC}&=-2\ln L_0+2p\,,\\
{\rm BIC}&=-2\ln L_0+p\ln N\,.
\end{align}
\end{subequations}

\begin{figure*}
\centering
\includegraphics[width=0.97\hsize,clip]{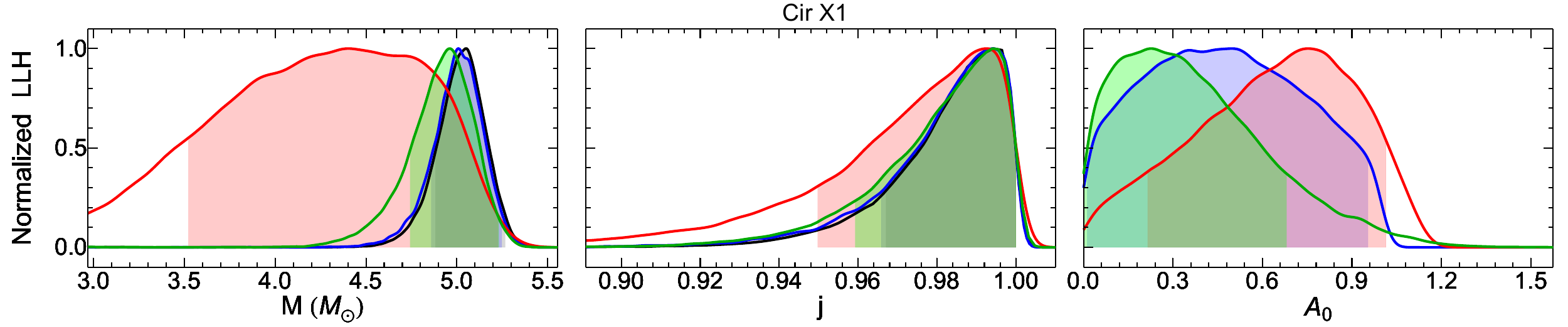}
\includegraphics[width=0.97\hsize,clip]{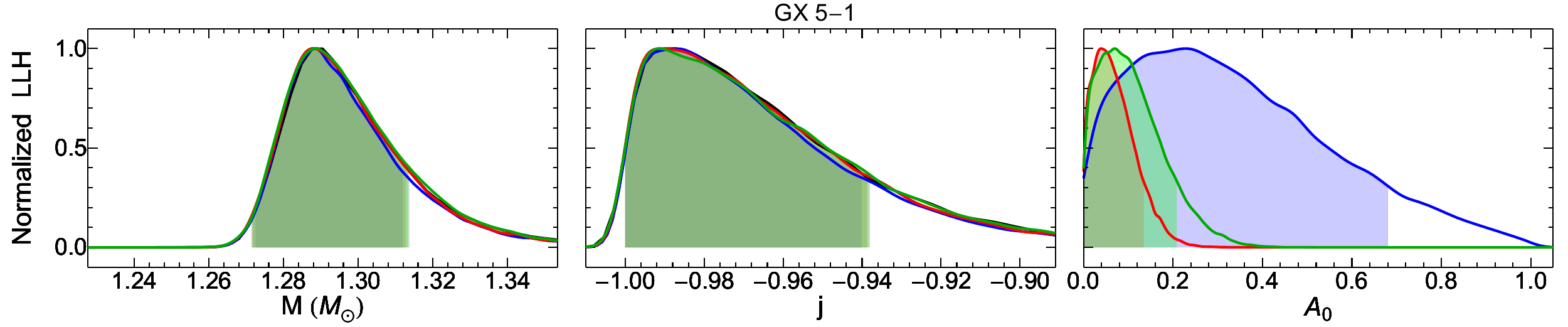}
\includegraphics[width=0.97\hsize,clip]{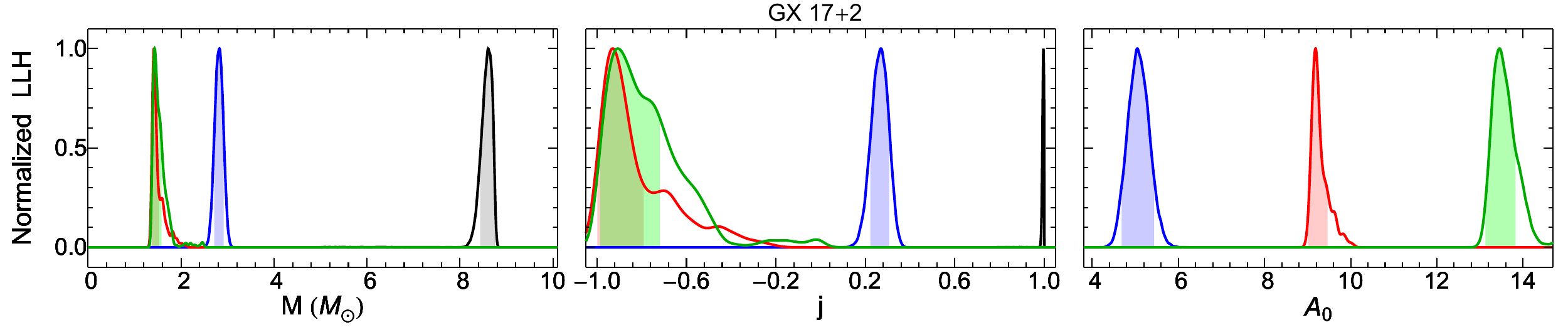}
\includegraphics[width=0.97\hsize,clip]{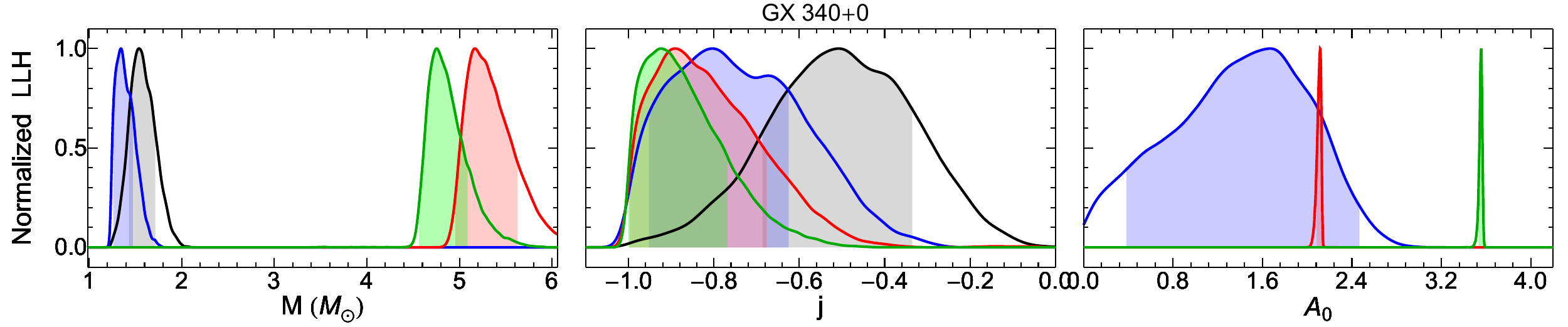}
\caption{Normalized $1$--D LLH functions (solid lines) and the $1$--$\sigma$ errors (shaded areas) for K spacetime within the harmonic case (black) and the anharmonic corrections A$_1$ (blue), A$_2$ (red), ad A$_3$ (green) applied to each of the sources listed in Tab.~\ref{tab:results_MCMC}.}
\label{fig:contours}
\end{figure*}

Next, the fiducial (best-suited) model is identified as the one with the lowest $\{{\rm AIC},{\rm BIC}\}$, say $\{{\rm AIC},{\rm BIC}\}_0$.

Finally, for each model, we calculate $\{\Delta{\rm AIC},\Delta{\rm BIC}\} = \{{\rm AIC},{\rm BIC}\} - \{{\rm AIC},{\rm BIC}\}_0$ and deduce weak, mild or strong statistical evidence in favor of the fiducial model if $\{\Delta{\rm AIC},\Delta{\rm BIC}\}\in[0,3]$, $\{\Delta{\rm AIC},\Delta{\rm BIC}\}\in (3,6]$, or $\{\Delta{\rm AIC},\Delta{\rm BIC}\}>6$, respectively.

\begin{figure*}
\centering
\includegraphics[width=0.97\hsize,clip]{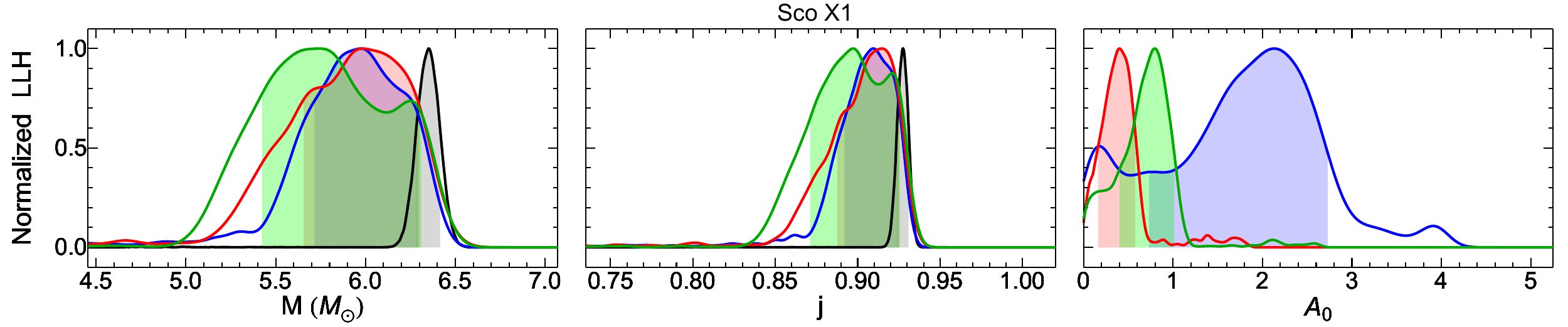}
\includegraphics[width=0.97\hsize,clip]{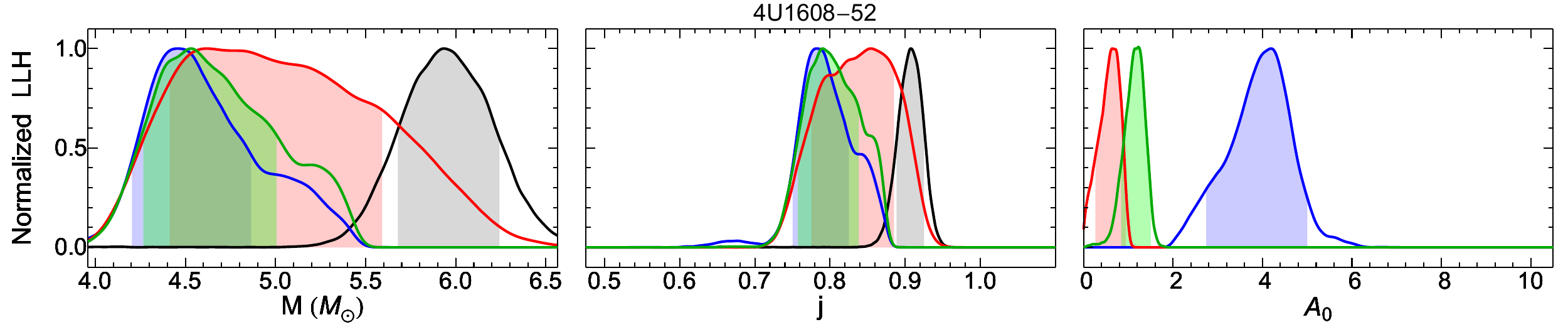}
\includegraphics[width=0.97\hsize,clip]{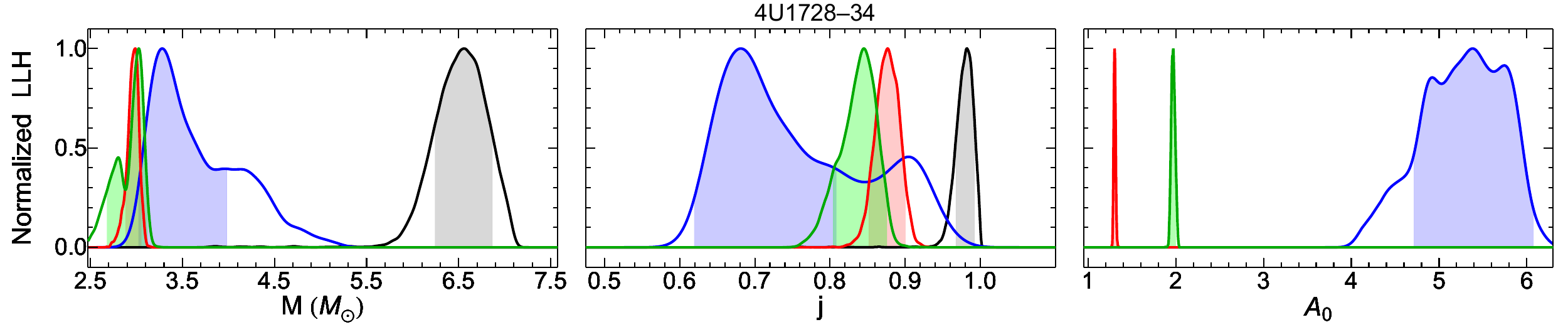}
\includegraphics[width=0.97\hsize,clip]{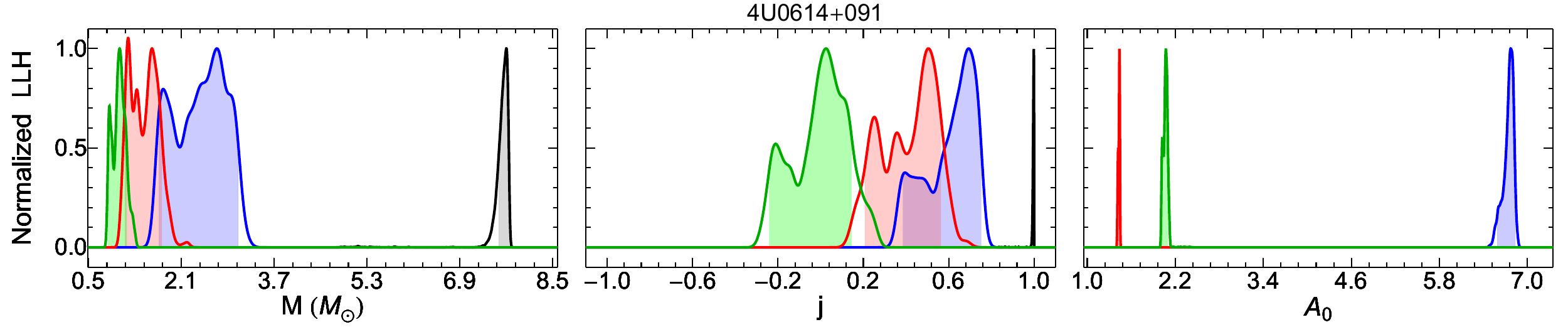}
\caption{Continued Fig.~\ref{fig:contours}.}
\label{fig:contours2}
\end{figure*}

\begin{figure*}
{\hfill
\includegraphics[width=0.48\hsize,clip]{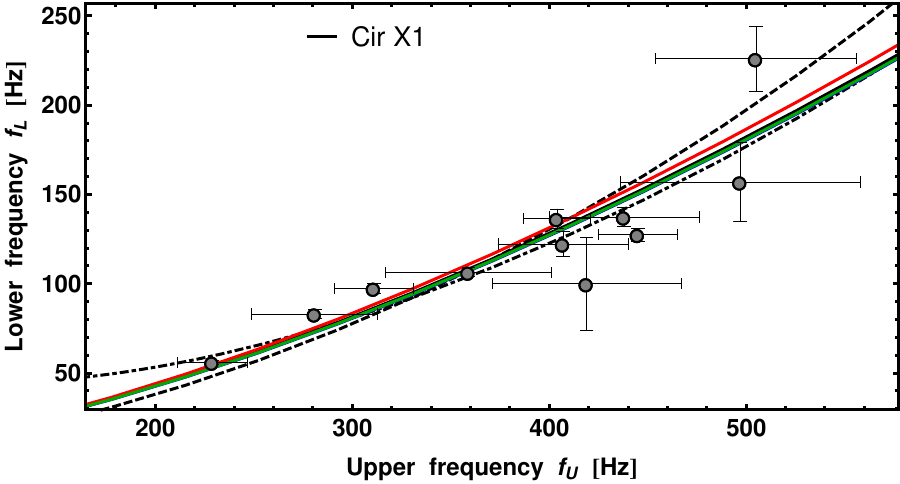}\hfill
\includegraphics[width=0.48\hsize,clip]{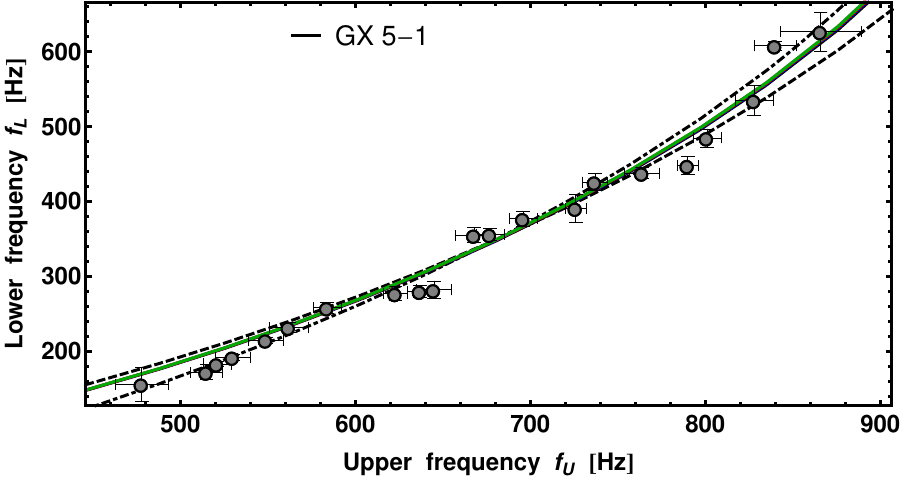}\hfill}

{\hfill
\includegraphics[width=0.48\hsize,clip]{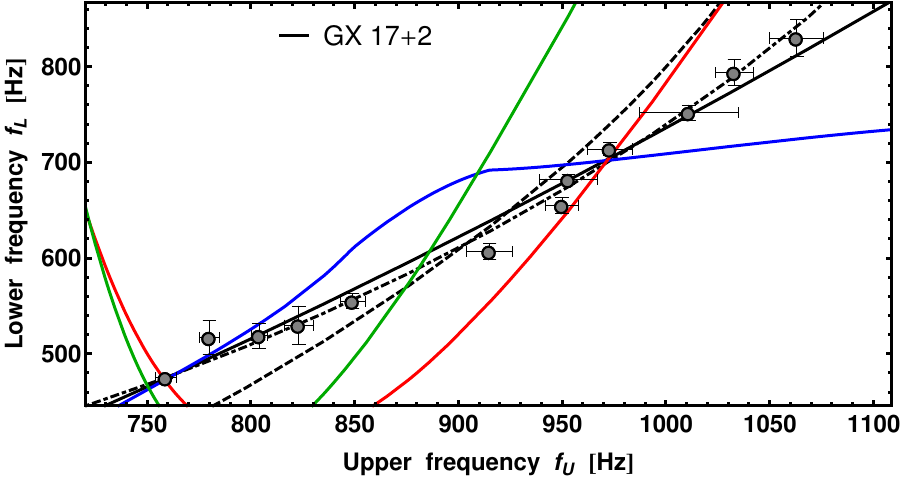}\hfill
\includegraphics[width=0.48\hsize,clip]{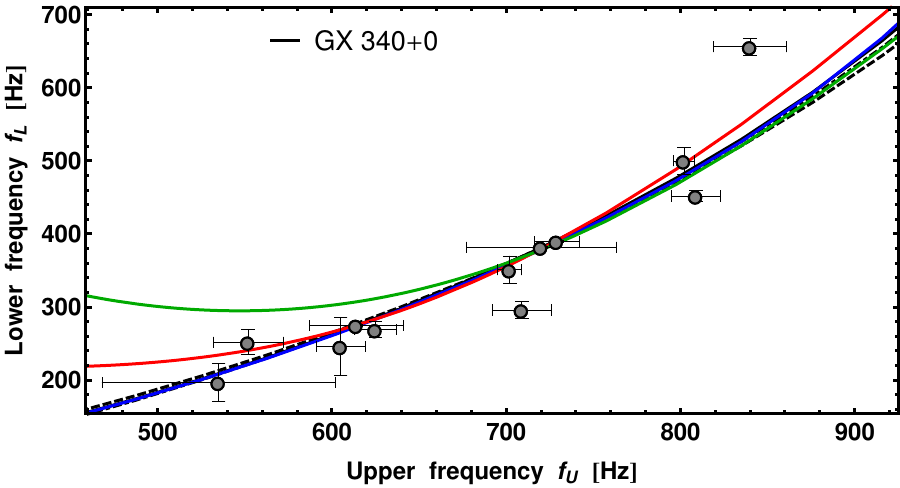}\hfill}

{\hfill
\includegraphics[width=0.48\hsize,clip]{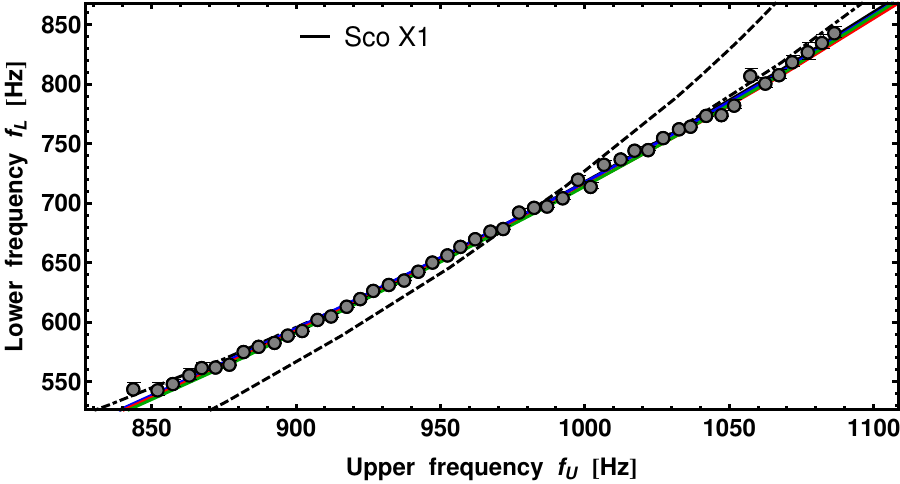}\hfill
\includegraphics[width=0.48\hsize,clip]{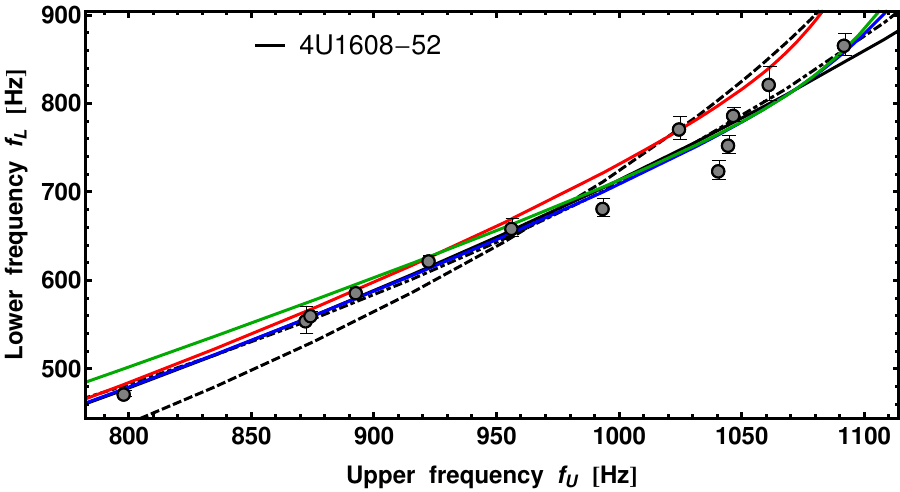}\hfill}

{\hfill
\includegraphics[width=0.48\hsize,clip]{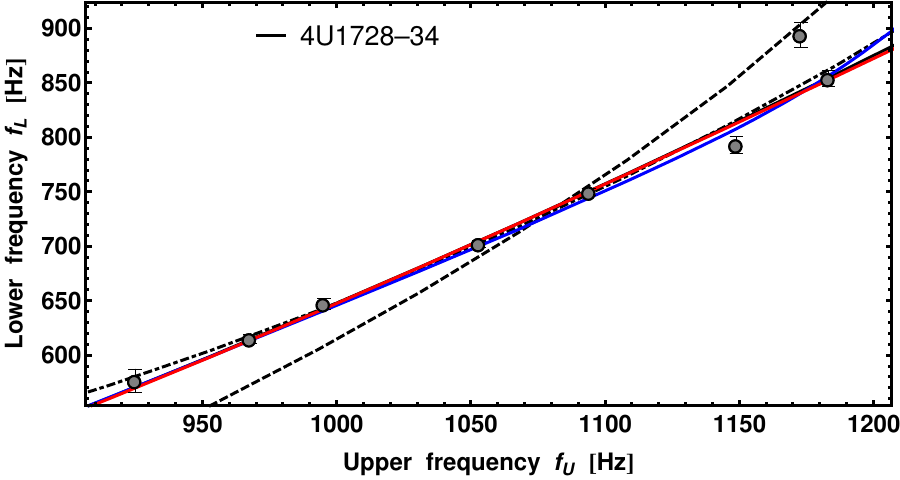}\hfill
\includegraphics[width=0.48\hsize,clip]{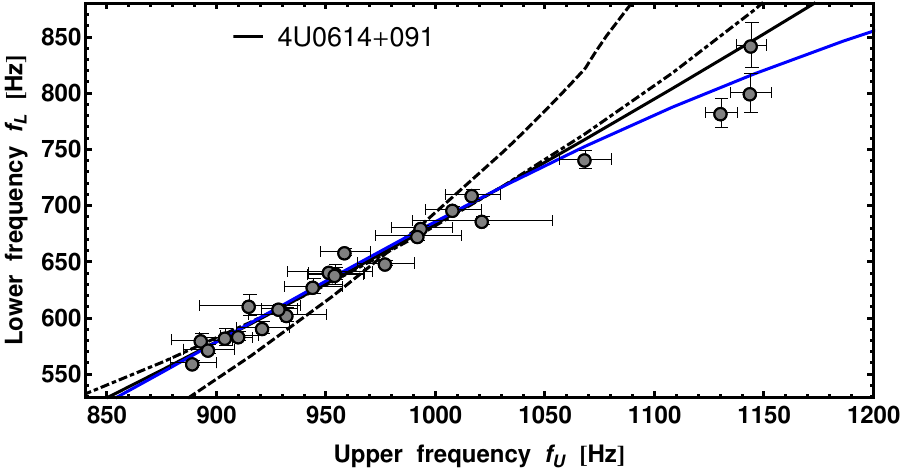}\hfill}
\caption{Plots of $f_{\rm L}$--$f_{\rm U}$ data sets and best-fitting curves: for the harmonic case S (dashed black), SdS (dot-dashed black), K (solid black) spacetimes, and for the anharmonic cases A$_1$ (solid blue), A$_2$ (solid red), and A$_3$ (solid green) applied to K metric.}
\label{fig:freq}
\end{figure*}

\subsection{Physical selection criteria}

Besides the statistical selection criteria, each specific model shall pass the final scrutiny of the NS physics.

The theoretical mass of a NS and its internal structure depend upon the selected equation of state (EoS). Moreover, theoretical expectations have to match the observational constraints.
Therefore, in drawing our conclusions, we adopt the physical selection criteria listed below.
\begin{itemize}
\item[-] \textit{Mass}. Depending on the EoS, NSs can have a maximum theoretical mass of 3.2 $M_\odot$ \citep{2002BASI...30..523S}.
\item[-] \textit{Spin parameter}. For a wide class of realistic EoS, a NS can reach up to $j \sim 0.7$ \cite{2011ApJ...728...12L}. Though crust-less NSs could in principle achieve $j\in(0.7,1]$ \cite{2016RAA....16...60Q}, we do not consider such exotic configurations.
\end{itemize}

\subsection{Summary of the analyses}

Our results are reported in Table~\ref{tab:results_MCMC}.
The normalized ($\ln L/\ln L_0$) one-dimensional ($1$--D) LLH functions for K spacetimes, in both the harmonic case and the anharmonic models, are displayed in Figs.~\ref{fig:contours}--\ref{fig:contours2}.
Moreover, for the sake of completeness, the fits of the QPO lower $f_{\rm L}$ and upper $f_{\rm U}$ frequencies for each model, in both harmonic and anharmonic cases, are shown in Fig.~\ref{fig:freq}.

Below we summarize our results for each QPO source.
\begin{itemize}
\item[-] {\bf Cir~X-1} \cite{2006ApJ...653.1435B}. Within the harmonic approximation, the SdS spacetime is the best-fitting scenario \cite{2023PhRvD.108l4034B}. The K spacetime is also physically ruled out, in both harmonic and anharmonic frameworks, being masses and spin parameters incompatible with the NS interpretation \cite{2024MNRAS.531.3876B}.
\item[-] {\bf GX~5-1} \cite{1998ApJ...504L..35W,2002MNRAS.333..665J}.
Also in this case, the SdS metric in the harmonic approximation is the best-fitting scenario \cite{2023PhRvD.108l4034B}. K spacetime fails in being predictive because of the high values of $j$ \cite{2024MNRAS.531.3876B}.
\item[-] {\bf GX~17+2} \cite{2002ApJ...568..878H}.
The harmonic scenario with SdS metric is hereby confirmed \cite{2023PhRvD.108l4034B}. Although in the anharmonic scenarios the high-mass issue of the K spacetime seems to be alleviated (and the spin parameter of the A$_1$ case falls into NS expectations), these approximations completely fail in reproducing the $f_{\rm L}$--$f_{\rm U}$ behavior, as showcased in Fig.~\ref{fig:freq}.
\item[-] {\bf GX~340+0} \cite{2000ApJ...537..374J}. Statistically, the best-fit scenario here is the A$_2$ case, which is however physically excluded in view of the high mass and spin. The A$_1$ model parameters, within the errors, are in agreement with NS physics. However, this anharmonic model is only weakly (mildly) favored with respect to the harmonic scenario of the K (SdS) spacetime.
\item[-] {\bf Sco~X1} \cite{2000MNRAS.318..938M}. The harmonic case and all anharmonic models applied to K spacetime are, respectively, the best-fitting scenario and good alternatives (as they are weakly/mildly excluded), albeit physically rejected in view of the high values of $M$ and $j$. In the harmonic case, the SdS spacetime seems to provide physically supported results, with a mass close to the range $1.40$--$1.52$~M$_\odot$ inferred from the optical light curves modelling \cite{2021MNRAS.508.1389C}.
\item[-] {\bf 4U1608-52} \cite{1998ApJ...505L..23M}. Also here, K metric in both harmonic and anharmonic scenarios is statistically favored, but physically rejected because of the high values of $M$ and/or $j$. Again, SdS spacetime in the harmonic approximation is the viable scenario.
\item[-] {\bf 4U1728-34} \cite{1999ApJ...517L..51M}. The SdS metric in the harmonic approximation is statistically and physically suitable \cite{2023PhRvD.108l4034B}. Viable scenarios, with parameters conforming NS physics, are also A$_1$ and A$_2$ cases (weakly and mildly excluded, respectively).
\item[-] {\bf 4U0614+091} \cite{1997ApJ...486L..47F}. The anharmonic model A$_1$ is statistically and physically suitable. Albeit with largely different values of $M$ and $j$, the A$_3$ model is a weakly excluded, but still viable scenario.
\end{itemize}

\begin{table}
\centering
\footnotesize
\setlength{\tabcolsep}{1.5em}
\renewcommand{\arraystretch}{1.2}
\begin{tabular}{lll}
\hline\hline
Source                                  &
Best scenario                           &
Viable scenarios                        \\
\hline
Cir X1                                  &
H--SdS                                  &
--                                      \\
GX 5-1                                  &
H--SdS                                  &
--                                      \\
GX 17+2                                 &
H--SdS                                  &
--                                      \\
GX 340+0                                &
A$_1$--K                                &
H--K, H--SdS                            \\
Sco X1                                  &
H--SdS                                  &
--                                      \\
4U1608–52                               &
H--SdS                                  &
--                                      \\
4U1728–34                               &
H--SdS                                  &
A$_1$--K, A$_2$--K                      \\
4U0614+091                              &
A$_1$--K                                &
A$_3$--K                                \\
\hline
\end{tabular}
\caption{Summary of the best-fitting and viable scenarios (approximation--spacetime) for each QPO source in this work.}
\label{tab:results_MCMC2}
\end{table}

Looking at Table~\ref{tab:results_MCMC2}, we draw some general conclusions.
\begin{itemize}
\item[-] When the harmonic approximation is considered, six out of eight sources are best-fit by SdS spacetime. Accordingly, one can deduce that the energy-momentum tensor shall account for a cosmological constant term that usually relates to quantum fluctuations of vacuum, implying some sort of quantum nature of NSs. However, this appears quite unphysical in view of the classical nature of the sources. In general, it seems evident that the Kerr hypothesis is not compatible with the RPM paradigm, limiting its robustness in describing QPOs.
\item[-] The anharmonic approximation seems to overcome the above issue in two sources out of eight, showing that \emph{QPOs cannot be handled at first order, but at least at second}, and that the best scenario is K spacetime. However, SdS is still a viable replacement for the anharmonic models and viceversa. Accordingly, the Kerr hypothesis is healed here and appears to be more suitable than the usual harmonic case, commonly adopted in the literature.
\item[-] In all the sources (excluding GX~17+2, where convergence was not always achieved, see Table~\ref{tab:results_MCMC} and Fig.~\ref{fig:freq}), the parameters $R_0$ and $j$ display concordant signs.
This feature seems to oddly relate quantum properties with classical ones, hinting that both spacetimes may not be the best-suited ones to describe QPO, or that RPM is still incomplete. In other words, to fully satisfy the Kerr hypothesis, anharmonic corrections look essential, as well as additional modifications of the standard RPM treatment which therefore needs evident refinements, being, however, beyond the scope of current work.
\end{itemize}

\section{Final remarks and perspectives}
\label{final}

In this work, we updated the RPM paradigm of QPOs by including higher-order corrections to the geodesic motion. In particular, we explored the impact of anharmonic terms arising from the third-order Taylor expansion of the effective potential. We therefore critically revised the standard RPM paradigm, including the Kerr hypothesis inside it. In this respect, we first demonstrated that the aforementioned hypothesis is not recovered inside the RPM standard recipe and, moreover, that linear order approximation appears incomplete, in fitting QPO data points.

Our analysis demonstrated that, regardless of the background spacetime geometry, the only relevant anharmonic correction appears in the radial sector, and is quadratic in the radial perturbation $\delta r$. This term contributes \emph{a non-negligible correction to the radial frequency}. This appears quite more evident, especially near the ISCOs, indicating that the linear order cannot be definitive to describe QPOs.
Conversely, polar perturbations $\delta \theta$ remained effectively decoupled from the radial dynamics, confirming the approximate separability of oscillation modes in the perturbative regime.

To this end, we thus performed MCMC analyses on eight NS sources of QPOs, comparing model predictions with observational data. Accordingly, we reviewed the standard harmonic approach in S, SdS, and K spacetimes, confirming previous literature that tended to favor the presence of a de Sitter phase to fit QPO data points, albeit limiting to the harmonic approximation.

Subsequently, we included the leading anharmonic correction in the K geometry and reassessed the model fits. Even though the inclusion of anharmonic terms significantly improved the internal consistency of the epicyclic frequency expansion, discrepancies between theoretical predictions and observed QPO patterns did not disappear.
The angular momenta and overall masses, obtained by bounding the K metric with the eight sources did not match theoretical expectations, showing inconsistent results that cannot support the robustness of the RPM paradigm.
More precisely, these results indicated that, while anharmonic corrections are physically motivated and statistically relevant, they do not fully resolve the observed tensions between the harmonic RPM-based models and data.

Our findings suggested that the standard picture of QPOs under the RPM framework must be generalized to include non-linear contributions that cannot be discarded \emph{a priori}, as commonly reported in the literature.

Even though these corrections appear essential, they turn out to be insufficient to provide a complete phenomenological description, opening new avenues toward the way of approximating the geodesic equations to characterize QPOs.

In other words, we definitely showed that the standard RPM is not predictive and physical findings cannot be expected, using a naive harmonic approximation over $r$ and $\theta$, violating \emph{de facto} the Kerr hypothesis.

Further improvements may require the inclusion of additional physical mechanisms such as pressure gradients, magnetic fields, or fully relativistic fluid perturbations. Moreover, the role of spacetime deviations from the K metric, possibly induced by quantum-gravitational or exotic matter effects, requires investigation.
It is clearly possible, in fact, that the model itself could be generalized by including additional effects, such as external fields, rather than expanding to higher orders only or working alternative expansions out.

In so doing, perspectives will focus on a systematic extension of the model to account for mode coupling beyond the leading order, as well as the implementation of non-geodesic corrections arising in magneto-hydrodynamic or self-gravitating disk models. Additionally, comparing our results with those obtained in alternative QPO background theories may help in clarifying the compact object nature and properties. Last but not least, high-precision timing data from next-generation X-ray missions will be crucial in testing these theoretical developments. Finally, the RPM is assumed to be a viable scheme, as it directly comes from general relativity. Potential deviations from its use in framing QPO data may even indicate the need of extending gravity at a fundamental level. Although this prerogative is clearly the least likely, especially focusing on QPO data points, it would not have to be ignored, deserving additional speculation.

\section*{Acknowledgments}
OL acknowledges the support by the  Fondazione  ICSC, Spoke 3 Astrophysics and Cosmos Observations. National Recovery and Resilience Plan (Piano Nazionale di Ripresa e Resilienza, PNRR) Project ID $CN00000013$ ``Italian Research Center on  High-Performance Computing, Big Data and Quantum Computing" funded by MUR Missione 4 Componente 2 Investimento 1.4: Potenziamento strutture di ricerca e creazione di ``campioni nazionali di R\&S (M4C2-19)" - Next Generation EU (NGEU). MM acknowledges support from the project OASIS, ``PNRR Bando a Cascata da INAF M4C2 - INV. 1.4''.

\end{document}